\documentstyle[12pt,emulateapj,amssym,psfig]{article}

\newcommand\lea{\mathrel{\raise .4ex\hbox{\rlap{$<$}\lower 1.2ex\hbox{$\sim$}}}}
\newcommand\gea{\mathrel{\raise .4ex\hbox{\rlap{$>$}\lower 1.2ex\hbox{$\sim$}}}}
\newcommand\solar{\ifmmode_{\mathord\odot}\else$_{\mathord\odot}$\fi}

\newcommand\z{$\pm$}
\renewcommand\deg{\ifmmode^\circ\else$^\circ$\fi}

\received{March 8, 2000}
\accepted{August 25, 2000}

\slugcomment{To appear in {\it The Astrophysical Journal,} v.546, n.2}

\lefthead{Lobanov A.P. Gurvits L.I., Frey S., et al.}

\righthead{VSOP observation of the quasar PKS\,2215+020: a new 'core-jet'
laboratory at $z=3.57$}

\psdraft
\begin{document}

\title{VSOP observation of the quasar PKS\,2215+020: a new laboratory for
core-jet physics at $z=3.572$}

\author{A.~P.~Lobanov\altaffilmark{1},
L.~I.~Gurvits\altaffilmark{2,3}, S.~Frey\altaffilmark{4},
R.~T.~Schilizzi\altaffilmark{2,5},
K.~I.~Kellermann\altaffilmark{6}, N.~Kawaguchi\altaffilmark{7},
I.~I.~K.~Pauliny-Toth\altaffilmark{1}}

\altaffiltext{1}{Max-Planck-Institut f\"{ur} Radioastronomie, Auf dem
H\"{u}gel 69, D-53121 Bonn, Germany; alobanov@mpifr-bonn.mpg.de, 
ipauliny@mpifr-bonn.mpg.de}
\altaffiltext{2}{Joint Institute for VLBI in Europe, P.O.Box 2, 7990 AA, 
Dwingeloo, The Netherlands; lgurvits@jive.nl, schilizzi@jive.nl} 
\altaffiltext{3}{Astro Space Center of P.N.Lebedev Physical Institute, 
Moscow, Russia}
\altaffiltext{4}{F\"{O}MI Satellite Geodetic Observatory, P.O. Box 546, 
H-1373 Budapest, Hungary; frey@sgo.fomi.hu} 
\altaffiltext{5}{Leiden Observatory, P.O. Box 9513, 2300 RA, Leiden, 
The Netherlands}
\altaffiltext{6}{National Radio Astronomy Observatory, 520 Edgemont Rd, 
Charlottesville, VA 22903-2475, USA; kkellerm@nrao.edu} 
\altaffiltext{7}{NAO, Mitaka, Tokyo 181-8588, Japan,
kawagu@hotaka.mtk.nao.ac.jp}

\begin{abstract}

We report results of a VSOP (VLBI Space Observatory Programme)
observation of a high redshift quasar PKS\,2215+020 ($z=3.57$). The
$\sim$1 milliarcsecond resolution image of the quasar reveals a
prominent `core-jet' structure on linear scales from 5$h^{-1}$ to
300$h^{-1}$~pc ($H_{\circ}=100\,h$~km\,s$^{-1}$Mpc$^{-1}$). The
brightness temperatures and sizes of bright features identified in the
jet are consistent with emission from relativistic shocks dominated by
adiabatic energy losses.  The jet is powered by the central black hole
with estimated mass of $\sim 4 \times 10^{9}{\rm
M}_{\solar}$. Comparisons with VLA and ROSAT observations indicate a
possible presence of an extended radio/X--ray halo surrounding
2215+020.

\end{abstract}


\section{Introduction}

VLBI\footnote{Very Long Baseline Interferometry} observations of high
redshift quasars at centimeter wavelengths allow to address two
connected topics:

{\sl (1) Frequency-dependent properties of the milliarcsecond
structures in quasars.} An image obtained at a received frequency
$\nu_{\rm r}$ represents the source structure at a higher emitted
frequency $\nu_{em}=\nu_{obs}(1+z)$ thus facilitating a direct
comparison between radio structures at a given frequency in high
redshift quasars and their low redshift counterparts observed at 
a higher frequency $\nu_{obs}$. 

{\sl (2) Cosmological tests using the milliarcsecond structure of QSOs
as a ``standard'' object.} As shown in several recent studies, the
``apparent angular size -- redshift'' ($\theta-z$, \cite{kel93},
\cite{gur94}, \cite{gur+99}) and ``proper motion -- redshift''
($\mu-z$, \cite{vc94}, \cite{ver96}, \cite{kik+99}) relation on
milliarcsecond scales have a potential to provide meaningful
cosmological information. In this respect, sources at extremely high
redshifts are of special interest.

In this paper we present a VSOP (VLBI Space Observatory
Programme, \cite{hir+98}) observation of the high-redshift quasar
PKS\,2215+020. The observations have been conducted as part of a VSOP
study of extremely high redshift quasars. The complete results of 
this project will be presented elsewhere (\cite{gur+00}).

The VSOP is a Space VLBI (SVLBI) mission utilizing an orbiting
8--meter antenna deployed on the Japanese satellite HALCA\footnote{The
Highly Advanced Laboratory for Communication and Astronomy}
(\cite{hir+98}) and the worldwide array of radio telescopes.  The
satellite has an elliptical orbit, with the apogee and perigee heights
at $\simeq$~21\,000\,km and $\simeq$~560\,km, respectively, and
orbiting period of roughly 6 hours. In each observation, data from the
satellite are recorded by a network of up to 5 tracking stations, and
subsequently correlated with the data from participating ground radio
telescopes.  Regular VSOP observations started in September 1997 at
1.6 and 5\,GHz.  A 1.6\,GHz VSOP observation of PKS\,2215+020
described here was made on September 14-15, 1997.

PKS\,2215+020 is an optically faint ($m_{\rm B} = 21.97$), radio-loud
quasar ($S_{\rm 5\,GHz} = 0.5$-0.6\,Jy; \cite{gw93}, \cite{gre+96})
included in the Parkes half-jansky flat-spectrum sample
(\cite{dri+97}). It has an emission redshift, $z=3.572$ (\cite{dri+97}); a
corresponding linear scale of 3.38\,$h^{-1}$\,pc\,mas$^{-1}$ (for the
Hubble constant $H_0=100\,h$\,km\,s$^{-1}$Mpc$^{-1}$ and
deceleration parameter $q_0=0.5$). The radio spectral index of
2215+020 is $\alpha^{5\,{\rm GHz}}_{2.7\,{\rm GHz}} =-0.15$ ($S_{\nu}
\propto \nu^{\alpha}$). Siebert \& Brinkmann (1998) have observed X-ray
emission from 2215+020 with the ROSAT High Resolution Imager (HRI), and
found only marginal evidence for possible elongation along the ${\rm
P.A.} = 60$-$70\deg$.

\placetable{tb:observation}

\section{VSOP observation and data reduction}

We observed the quasar PKS\,2215+020 at 1.6\,GHz, using the HALCA
satellite and a global VLBI array of 15 ground-based radio telescopes
(Table~\ref{tb:observation}). The observation lasted for 10 hours,
with HALCA data recorded for 7 hours. The data were recorded in the
VLBA format (\cite{rog95}), with a total observing bandwidth of
32\,MHz divided in two intermediate frequency (IF) bands, each having
256 spectral channels.  The tracking stations in Green Bank and
Goldstone (USA) were used for the HALCA data acquisition. The data
were correlated at the VLBA correlator in Socorro (\cite{ben95}), with
a pre--averaging time of 1.966 and 0.524\,sec for the
ground--ground and space--ground baselines respectively.  Fringe
visibilities were detected in the HALCA data recorded at both tracking
stations. The resulting sampling function ({\it uv}-coverage) of the
final correlated dataset is shown in Figure~\ref{fig1}, indicating an
improvement of resolution by a factor of $\sim$2.5 compared to
ground--based VLBI observations at the same frequency, owing to the 
ground--space baselines.

\placefigure{fig1}

We used AIPS\footnote{The NRAO Astronomical Image Processing System}
and the DIFMAP program (\cite{she+94}) for post-processing of the
correlated dataset.  The a priori amplitude calibration was determined
from the antenna gain and system temperature measurements, including
that of the HALCA radio telescope (\cite{mol+00}). After inspecting
the IF bandpasses, the last 46 channels in each IF were deleted, owing
to a cutoff in the filter bandpass. This has reduced the effective
observing bandwidth to 26.2\,MHz. We corrected the residual delays and
rates, using single-band (SB) and multi-band (MB) fringe fitting
(\cite{cot95}) with solution intervals of 5 minutes, and accepted all
solutions with ${\rm SNR} \ge 7$. To improve the SNR on the
space--ground baseline solutions, we fringe fitted the HALCA data, using
the source model obtained by imaging the ground--ground VLBI
data. After the fringe fitting, the residual phase variations on the
ground--ground baselines were within 3$\deg$, and within $15\deg$, on
the space--ground baselines. We then averaged over all frequency
channels, and calibrated the phases with a point source model (to
enable time averaging). Finally, the data were exported into DIFMAP and
further time--averaged into 60--second bins. The amplitude and phase
errors were re-calculated in DIFMAP, from the scatter in the unaveraged
data. The estimated RMS noise on the HALCA baselines was about 4 times
higher than on the ground--ground baselines, consistent with the
smaller antenna size and higher system temperature of the HALCA radio
telescope.  The amplitude visibility distribution obtained after the
amplitude calibration and fringe--fitting is shown in
Figure~\ref{fig2}.

\placefigure{fig2}

\section{Ground VLBI array and VSOP images of 2215+020}

From the final fringe--fitted dataset, we produced two images:  
a)~an image with all data included (VSOP image, hereafter), 
and b)~an image using only the ground--ground baseline data 
included (GVLBI image, hereafter). In both cases, uniform weighting 
is used, and gridding weights are scaled inversely with the amplitude 
errors. The resulting images are shown in Figure~\ref{fig3}, and 
their basic characteristics are presented in Table~\ref{tb:images}.

\placefigure{fig3}

\placetable{tb:images}

The images in Figure~\ref{fig3} show a clear core--jet type
morphology.  We note that the general milliarcsecond-scale morphology
of PKS\,2215+020 described here is consistent with the VLBA 15 GHz
image (\cite{kik+00}).  In the GVLBI image, we identify the core, C,
and 7 enhanced emission regions (jet components J1--J7, hereafter)
which we will discuss below.  The jet extends up to almost 80\,mas
from the core, and shows a particularly wide, bright section between
45 and 60\,mas (150--200\,$h^{-1}$pc projected distance), suggesting a
possible working surface of a young jet propagating in a dense ambient
medium.

The basic structures appear to be similar, in both images shown in
Figure~\ref{fig3}, taking into account the larger noise and
scatter of the data on the space--ground baselines, which results in
the increased RMS noise and stronger deconvolution effects in the VSOP
image. These effects can be countered by using the natural weighting
for the VSOP dataset. We find that the resolution and sensitivity of
an image of 2215+020 obtained from the naturally weighted VSOP data
are practically identical to those of the uniformly weighted ground
VLBI data from our observation (which is commonly expected for VSOP
observations; \cite{mur+99}). Since our goal is to study the jet
at highest resolution, we will discuss exclusively the uniformly
weighted data represented by the images in Figure~\ref{fig3}.

The most interesting difference between the two images shown in
Figure~\ref{fig3} is the absence of the component J4 in the VSOP
image. The adjacent
features J3 and J5 can be identified in both images, even though, in
the GVLBI image, they appear to be weaker than J4. It is not immediately
clear whether this effect results from unidentified imaging problems, or 
reflects in fact the true nature of J4 (which may be completely resolved 
in the VSOP image). 
Deconvolution effects are visible elsewhere in the two images (for
instance, in the VSOP image, J6 appears being split into two features). 
On the other hand, deconvolution alone is not likely to completely eliminate 
a feature. In order to provide quantitative criteria for deciding 
upon this issue, we turn now to modeling the source structure using the 
interferometric visibility data (\cite{pea95}). 

\placetable{tb:modelfit}

\subsection{Model fitting the jet structure in 2215+020}

To model the jet structure in 2215+020, we first use DIFMAP to fit the
ground baseline data by 8 circular Gaussian components (marked in
Figure~\ref{fig3}). Hereafter, we shall call this model ``the GVLBI
model''. The GVLBI model is then used as a starting guess for modeling
the visibilities in the entire VSOP dataset.  The resulting model
(``the VSOP model,'' hereafter) contains only 7 of the original 8
components. The component corresponding to feature J4 is indeed
completely resolved in the VSOP data.  In the VSOP model, the formal
size of J4 exceeds 20\,mas, with the flux of the feature being similar
to that in the VLBI model. We take this as evidence in favor of our
 suggestion that J4 is completely resolved in the VSOP data.

The derived component parameters are listed in
Table~\ref{tb:modelfit}, for both the GVLBI and VSOP models. To
estimate the uncertainties of the fitted component parameters, we
measure, for each feature, the peak flux density, $S_{\rm peak}$ and its 
RMS, in the
respective image. We then calculate the uncertainties using
approximations given by Fomalont (1989). 

In several features, both in the GVLBI and VSOP data, the fitted sizes
appear to be smaller than the resolution limit. In order to determine
which of the fitted values should be considered only as upper limit, we
derive, for each feature, its smallest detectable size of a circular 
Gaussian component in an image
with RMS noise $\sigma_{\rm RMS}$:
\begin{equation}
\label{eq:angsize}
d_{\rm lim} = \frac{\pi}{4} \left[ \frac{b_{\rm maj} b_{\rm min}}
{\ln(2) \ln(S/\sigma_{\rm RMS})} \right]^{1/2}\, ,
\end{equation}
where $S$ is the flux density of the feature, and $b_{\rm maj,min}$ are the 
major and minor axes of the restoring beam, respectively. For those features in
which $d<d_{\rm lim}$, we take $d_{\rm lim}$ as the upper limit
estimate of the size. 

Using the calculated flux densities and sizes, we calculate brightness
temperatures, $T_{\rm b}$, for all jet features.\footnote{We note that
our choice of Gaussian components for fitting the visibility data (as
opposed to, e.g., optically thick spheroids) does not
introduce qualitative changes to the results obtained. Models with
non--Gaussian brightness distribution would require only a small,
constant correction factor to be applied to the values of brightness
temperature presented in this paper.} For most of the calculated
$T_{\rm b}$, formal errors are of the order of 100\% (perhaps
indicating that the measured RMS errors of $S_{\rm peak}$ are too
conservative). We therefore regard the derived values of $T_{\rm b}$
as first order estimates. The component sizes and calculated
brightness temperatures are plotted in Figure~\ref{fig4}.

\placefigure{fig4}

\section{Discussion}

The VSOP observation shows several very interesting properties of the
jet in 2215+020. The observed extent of the jet ($>80\,{\rm mas},
250\,h^{-1}$\,pc) is remarkable. The structures observed in 2215+020
are almost ten times larger than in any other quasars at $z>3$
observed with VLBI (c.f. \cite{par+99} and references therein).  The
morphology of the source suggests that the component J1 may be a
working surface of the jet. From the jet expansion presented in
Figure~\ref{fig4}, we estimate that the jet opening angle should vary
between 2\deg and 7\deg (with the smaller value obtained not taking
into account the two most peculiar features: J4, and J1). These values
must also be corrected by a deprojection factor, $\sin\theta_{\rm
j}^{-1}$. However, considering the unusually large extent of the jet,
the effect of deprojection is not likely to be strong.

\subsection{Brightness temperature}

The observed brightness temperatures plotted in Figure~\ref{fig4} can
be used to probe the physics of the jet. Following Marscher (1990), we
assume that each of the jet components is an independent plane shock
in which the radio emission is dominated by adiabatic energy losses. The jet
plasma has a power law energy distribution, $N(E)d\,E \propto E^{-s}
d\,E$, and the magnetic field is assumed to vary as $B \propto
d^{-a}$. The Doppler factor is assumed to be constant or varying weakly
throughout the
jet. Under these assumptions, one can relate the brightness
temperature, $T_{\rm b,J}$, of each jet component to the brightness
temperature of the core, $T_{\rm b,C}$:
\begin{equation}
T_{\rm b,J} = T_{\rm b,C} \left( d_{\rm J}/d_{\rm C} \right)^{-\epsilon} \,,
\end{equation}
where $d$ denotes the respective measured sizes of the core and jet feature,
and $\epsilon = [2(2s+1)+3a(s+1)/6$. We take $s=2.0$ (corresponding to
synchrotron emission with a spectral index $\alpha=-0.5$) and $a=1$ (which
corresponds to the transverse orientation of magnetic field in the jet). The
measured value, $T_{\rm b,C}= 9\times 10^{11}$, is used for the core. We use
the sizes and upper limits measured in the VSOP image, and add the GVLBI
measurement for J4. The resulting brightness temperatures are indicated in
Figure~\ref{fig4} by thick dashed line.  The measured and model values of
brightness temperatures agree rather well, suggesting that the jet components
may indeed be a collection of relativistic shocks.  The discrepancy
  between the model and observed brightness temperatures is significant only
  in the outermost jet feature, J1, for which we find a ratio $\xi \equiv
  T_{\rm b}^{\rm obs}/T_{\rm b}^{\rm model} \approx 20$. This discrepancy
  suggests that J1 is a region of the jet where physical conditions may have
  become substantially different. If J1 interacts with the ambient medium, its
  material may be going through additional compression and
  reacceleration, which could lead to a localized increase of the brightness
  temperature. The observed discrepancy between our model and the measured
  brightness temperature of J1 may also be explained by variations of the jet
  Doppler factor, resulting from changes of the speed or orientation of the
  jet.  For this case, one can derive for the Doppler factor of J1,
  $\delta_{\rm J1} = \delta_{\rm j} \xi^{1/(\alpha-3)}$ (here $\delta_{\rm j}$
  is the Doppler factor in the rest of jet). This condition implies a jet
  bulk Lorentz factor $\Gamma_{\rm j} \ge \xi^{1/(3-\alpha)} = 2.5$ and
  viewing angle $\theta_{\rm jet} \ge \arccos[(\Gamma_{\rm j} -
  \xi^{1/(3-\alpha)})(\Gamma_{\rm j}^2 - 1)^{-1/2}]$. We find then that the
  Lorentz factor in J1 must be 40--50\% lower than in the rest of the jet
  ($\Gamma_{\rm J1}$ is lower than $\Gamma_{\rm j}$ because $\delta_{\rm j}
  <1$, so the jet has to be Doppler--deboosted, and its viewing angle must be
  $\ge 40^{\deg}$, even for $\Gamma_{\rm j}$ as high as 15). Alternatively, if
  the bulk Lorentz factor is constant throughout the entire jet, the direction
  of the flow entering the region of J1 must change by 20--30$^{\deg}$ with
  respect to the line of sight.

\subsection{The transverse size of the jet}

For the feature J4, which is resolved out in the VSOP image, the linear size
derived from the GVLBI data is $\sim 3.4$ mas ($12\,h^{-1}$ pc).  If this size
is related to the physical transverse dimension of the jet, the mass of the
central object can be estimated, assuming that the jet is collimated by the
ambient magnetic field of the host galaxy (e.g. Appl \& Camenzind 1992, 1993,
\cite{bm00}).  On parsec scales, the width of a jet can often be
underestimated, owing to limited resolution and dynamic range of VLBI
observations. In 2215+020, the jet feature J4 is arguably the best measure of
the width of the jet, as this feature is the only one resolved out in the VSOP
image, and its size (6.2\,mas) measured in the GVLBI image is significantly
larger than the respective limiting resolution (2.3\,mas). The measured sizes
of other jet features (except J1) are most likely affected by the limited
resolution and dynamic range of the data, and therefore are likely to be
smaller than the true jet width.  Finally, J1 is unsuitable for estimating the
central black hole mass because it is most likely a strongly perturbed region,
actively interacting with the external medium.

For a jet collimated by the ambient magnetic field, $B_{\rm ext}$, of
the host galaxy, the mass of the central object, $M_{\rm BH}$, can be
related to the width of the jet, $r_{\rm j}$, so that (\cite{bes97}):
\begin{eqnarray}
\label{eq:mass}
M_{\rm BH}& \sim 0.5\,r_{\rm j} c^2 G^{-1} (B_{\rm ext}/B_{\rm g})^{1/2} \approx \nonumber\\
& \approx r_{\rm j,pc} h^{-1} (B_{\rm ext}/B_{\rm g})^{1/2}10^{13}{\rm M}_{\solar}\, .
\end{eqnarray}
where $G$ is the gravitational constant, $R_{\rm g}$ is the Schwarzschild
radius of the the central black hole, and $B_{\rm g}$ is the magnetic field
measured at $R_{\rm g}$.  Equation (\ref{eq:mass}) refers to the
  transverse dimension of the jet measured at distances comparable to the
  collimation scale (typically expected to be located at 100--1000$R_{\rm g}$).
  
  A typical galactic magnetic field is $B_{\rm ext}\sim 10^{-5}$\,G
  (\cite{bec00}), and one can expect to have $B_{\rm g}\sim 10^4$\,G
  (\cite{fr93}).  With these estimates, we obtain $M_{\rm BH} \sim 4\times
  10^{9} h^{-1}{\rm M}_{\solar}$.  Since J4 is located at about 80 pc distance
  from the jet core, its size is most likely affected by the expansion of the
  jet.  This effect may result in overestimating the mass of the central black
  hole by $\lea30$\% (assuming a jet opening angle of $ 4^{\deg}$
  ).\footnote{Similar calculations yield $\sim 6\times 10^{8}{\rm M}_{\solar}$
    for the mass of the central black hole in M87, based on images from
    ground VLBI observations made at 1.6 GHz (\cite{rei+89}). This estimate
is comparable with the value of $\sim 3\times 10^9 {\rm M}_{\solar}$ deduced
from HST observations of M87 (\cite{har+94}). }

\placefigure{fig5}

\subsection{Large scale radio and X-ray structures in 2215+020}

The jet direction observed in our VLBI images is consistent with the
extension seen in a VLA observation (Figure~\ref{fig5}) on a scale of
7 arcseconds (23$h^{-1}$\,kpc). Most interestingly, a similar
extension is suggested by the ROSAT HRI image (ROSAT archive directory
701900h\footnote{ftp://ftp.xray.mpe.mpg.de/rosat/archive/700000/701900h/}).
Siebert \& Brinkmann (1998) argued that the extension of the emission
in the ROSAT image may result from the asymmetrical point--spread
function of the HRI.  In Figure \ref{fig6}, we overlay the ROSAT and
VLA images of 2215+020.  Given the remarkable morphological similarity
of the X--ray and radio emission, we tend to conclude that the
extension seen in the ROSAT may be real. This would imply that
2215+020 contains an X--ray emitting gas extending out to about
50$h^{-1}$\,kpc. The X-ray emission may then arise from interaction
between the relativistic outflow and a dense circumgalactic gas. It is
however not possible to say at present whether the emission on this
scales is directly related to the outflow observed on parsec scales.
Recent observations with Chandra X--Ray satellite (Marshall et
al. 2000) have indicated similar coincidence of radio and X--ray
emission distributions in several extragalactic radio sources, most
notably in NGC\,1275, allowing to speculate that radio lobes may be
pushing the X--ray emitting gas outwards. A similar scenario may be at
work in 2215+020.

\placefigure{fig6}

\section{Summary}

The main results from the observations reported in this paper can be
summarized as follows:

1.~The radio source PKS\,2215+020 has a rich core--jet morphology and
unusually large size of the jet which can be traced up to
300$h^{-1}$ parsecs. This is by far the longest jet observed at
redshifts above 3.

2.~The high resolving power of the VSOP allows us to measure the
transverse width of the jet, which provides a rough estimate for the
mass of the central black hole in the object $M_{\rm BH} \sim 4\times
10^{9} h^{-1}{\rm M}_{\solar}$.

3.~The sizes and brightness temperatures derived from the model of the
source structure are consistent with the adiabatic energy losses in 
relativistic shocks embedded in the jet.

4.~The arcsecond--scale structure detected in PKS\,2215+020 with the
  VLA coincides with the extension observed in the ROSAT X--ray
image, implying a possible presence of an X--ray emitting gas 
extending out to about 50$h^{-1}$ kiloparsecs.

The findings summarized above allow us to present the quasar 
PKS\,2215+020 as a promising laboratory for future studies of 
core--jet physics at very high redshifts.

\acknowledgments

We gratefully acknowledge the VSOP Project, which is led by the Institute of
Space and Astronautical Science (Japan) in cooperation with many agencies,
institutes, and observatories around the world.  We thank W.~Brinkmann and
M.~Gliozzi for their helpful remarks on the interpretation of the ROSAT data.
We would like to thank the anonymous referee for comments and suggestions for
improving the manuscript.  SF acknowledges support from the Hungarian Space
Office, NWO and OTKA.  LIG acknowledges partial support from the European
Commission TMR Programme, Access to Large--Scale Facilities under contract
ERBFMGEXT950012, the European Commission, TMR Programme, Research Network
Contract ERBFMRXCT~96--0034 ``CERES''. The National Radio Astronomy
Observatory is operated by Associated Universities, Inc.  under a Cooperative
Agreement with the National Science Foundation.


\clearpage


\figcaption[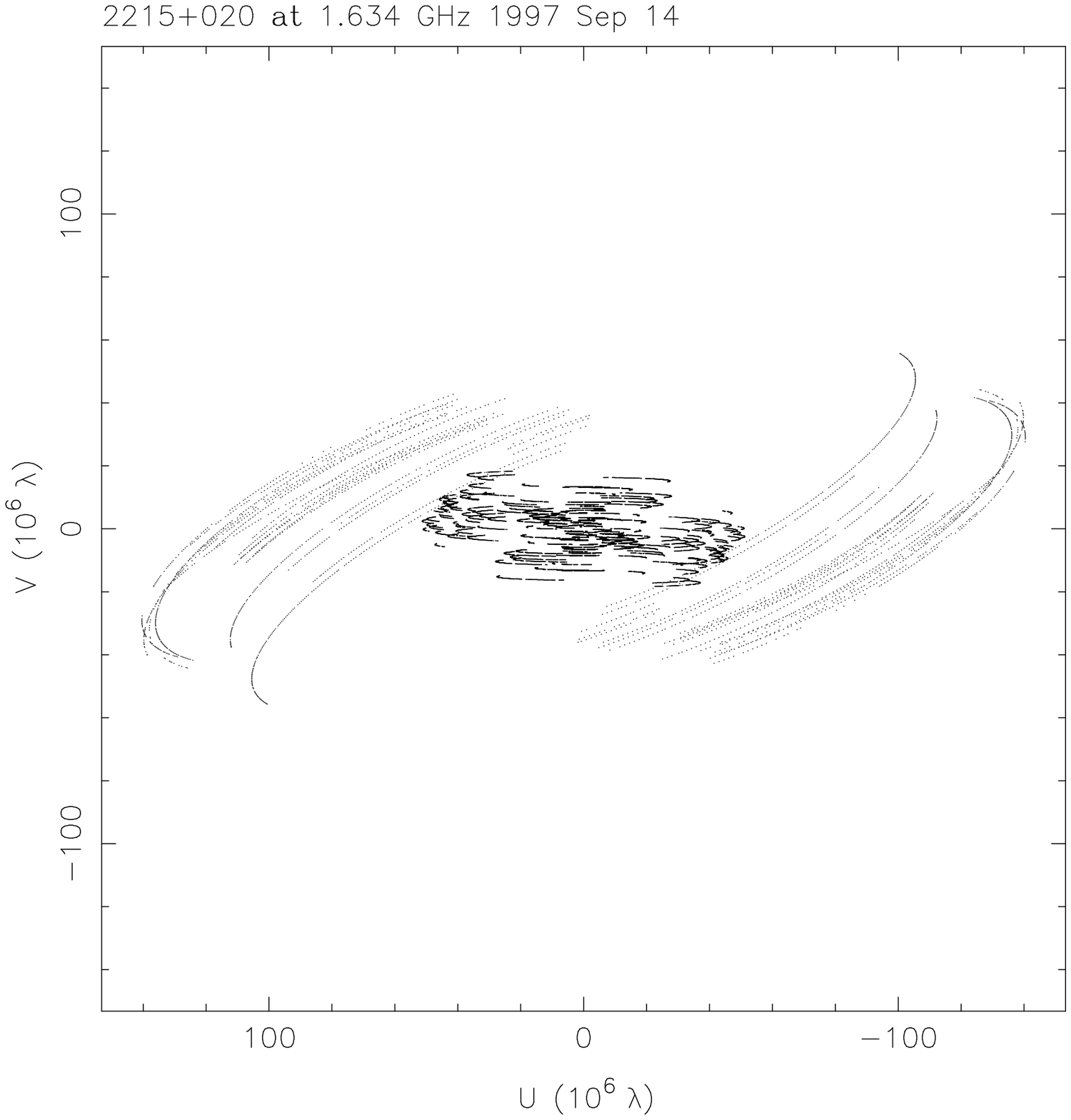]{The $(u,v)$-coverage of the observation of
2215+020. The largely horizontal, inner tracks are from the
ground--ground baselines; the inclined, outer sections are
from the ground--space baselines.
\label{fig1}}

\figcaption[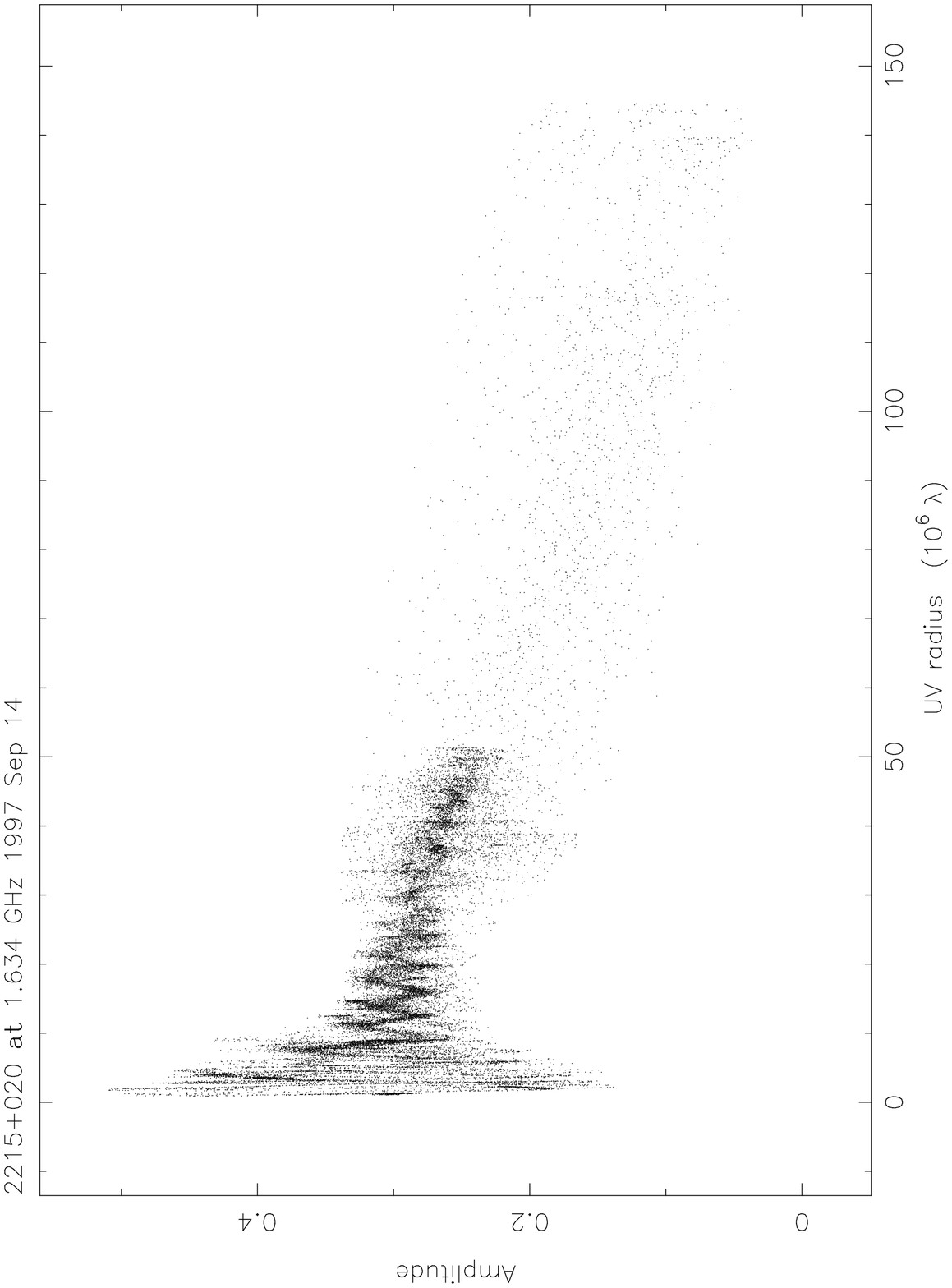]{The distribution of the correlated flux versus
$uv$--distance. 
The ground--ground baselines extend up to 50\,M$\lambda$; the ground--space
baselines cover the $uv$--distances between 20 and 150\,M$\lambda$.
\label{fig2}}

\figcaption[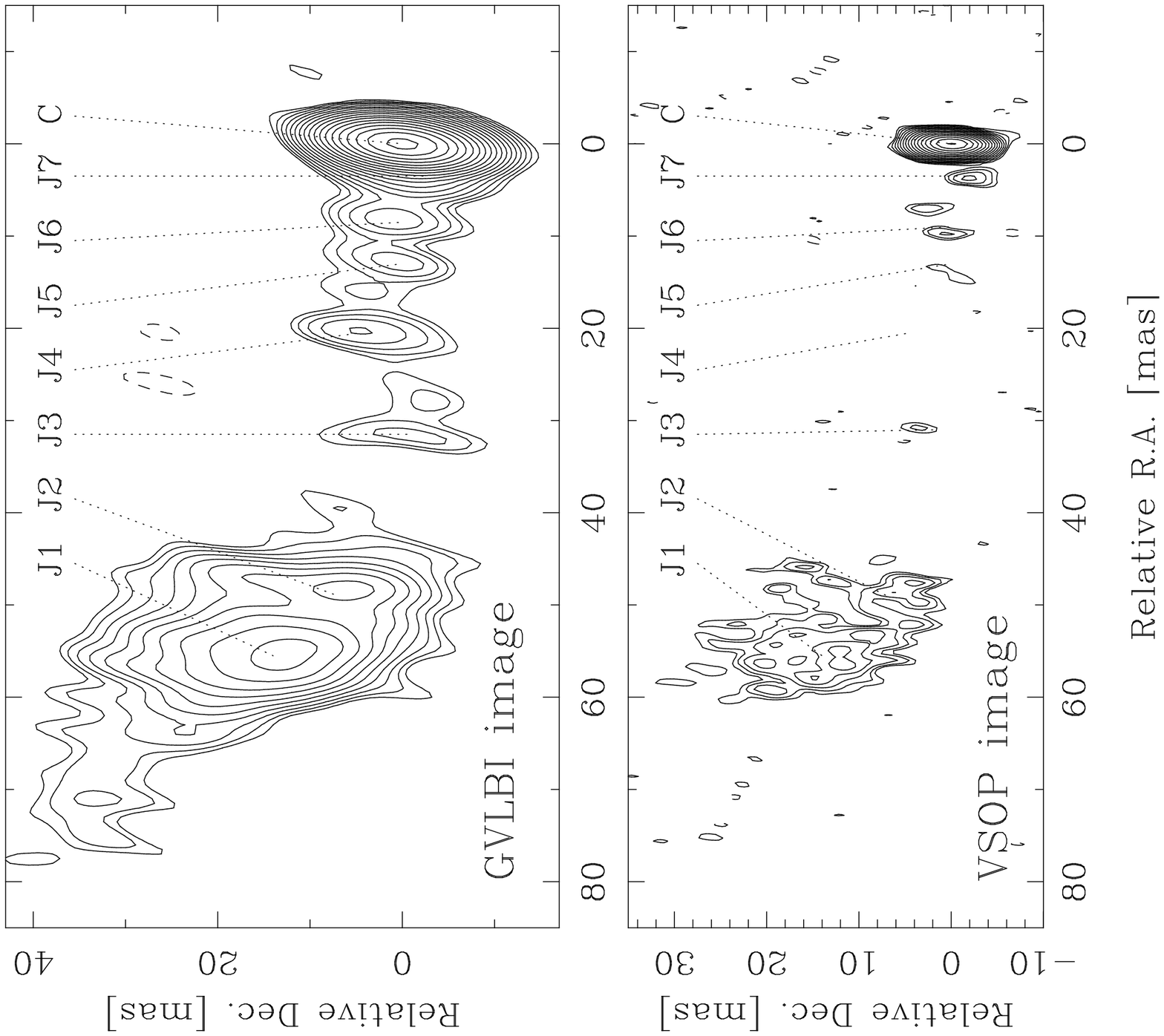]{The GVLBI (top) and VSOP (bottom) images
of 2215+020. Image characteristics are summarized in Table~2.  The
lowest positive contours represent the brightnesses of 0.7\,mJy/beam
(GVLBI image) and 1.2\,mJy/beam (VSOP image). Contour levels are drawn
at $-$1.0, 1.0, 1.41, 2.0, 2.82, 4,~... of the lowest positive
contours.  Shown in both images are locations of the Gaussian
components obtained from model fitting the structures in the GVLBI
image. The modelfits are described in section~3.1. In the VSOP image:
the component J4 is resolved out; the component J6 appears to
be double, likely due to deconvolution errors owing to the much larger noise
on the ground--space baselines.
\label{fig3}}

\figcaption[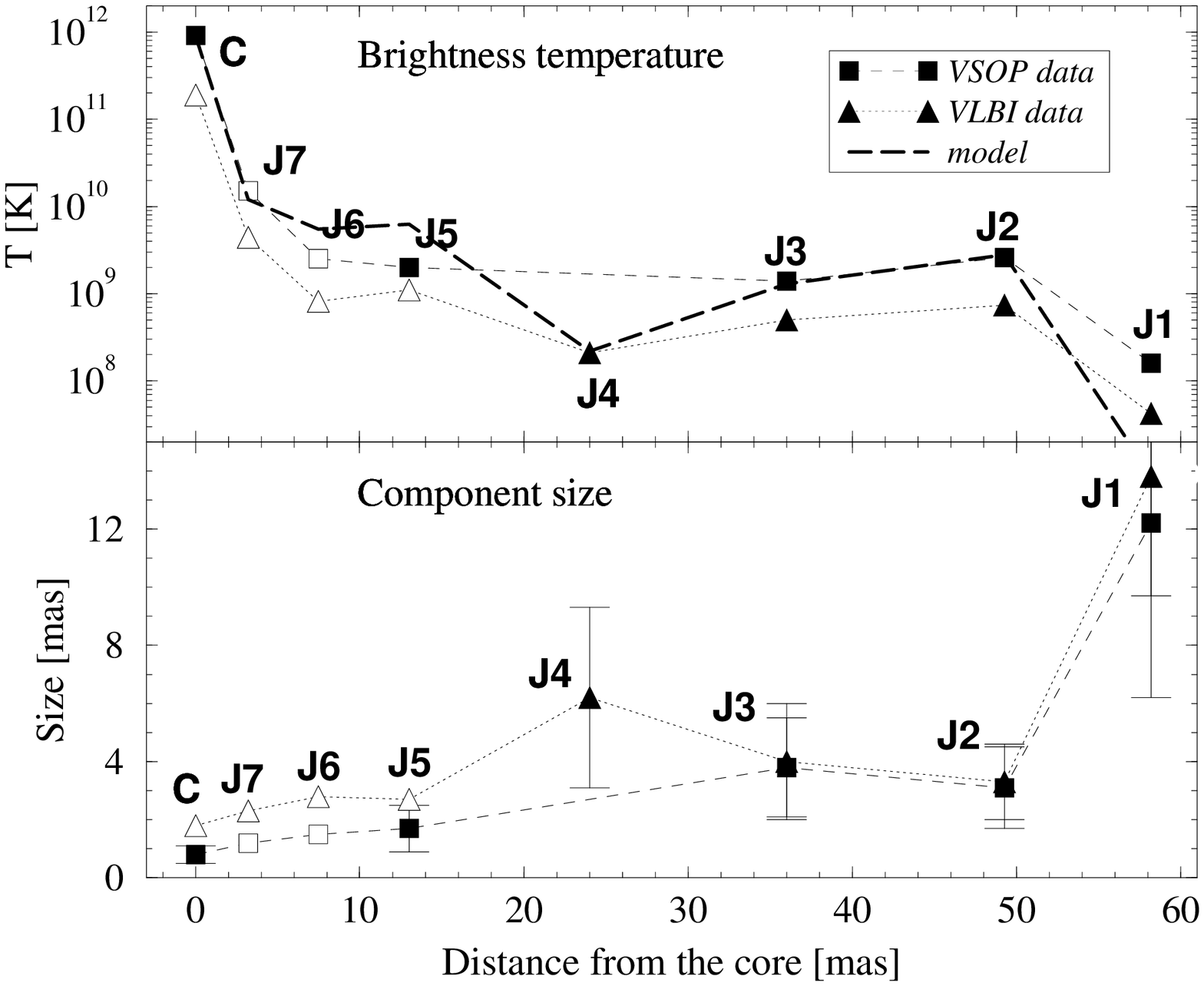]{Sizes and brightness temperatures of different
emitting components in the jet of 2215+020. The component
identification is described in Figure~3. Open symbols denote upper
limits of the size and lower limits of the brightness temperature. The
feature J4 is completely resolved in the VSOP image. In the brightness
temperature plot, the thick dashed line represents values predicted 
for  relativistic shocks in which adiabatic energy losses dominate
the radio emission.
\label{fig4}}

\figcaption[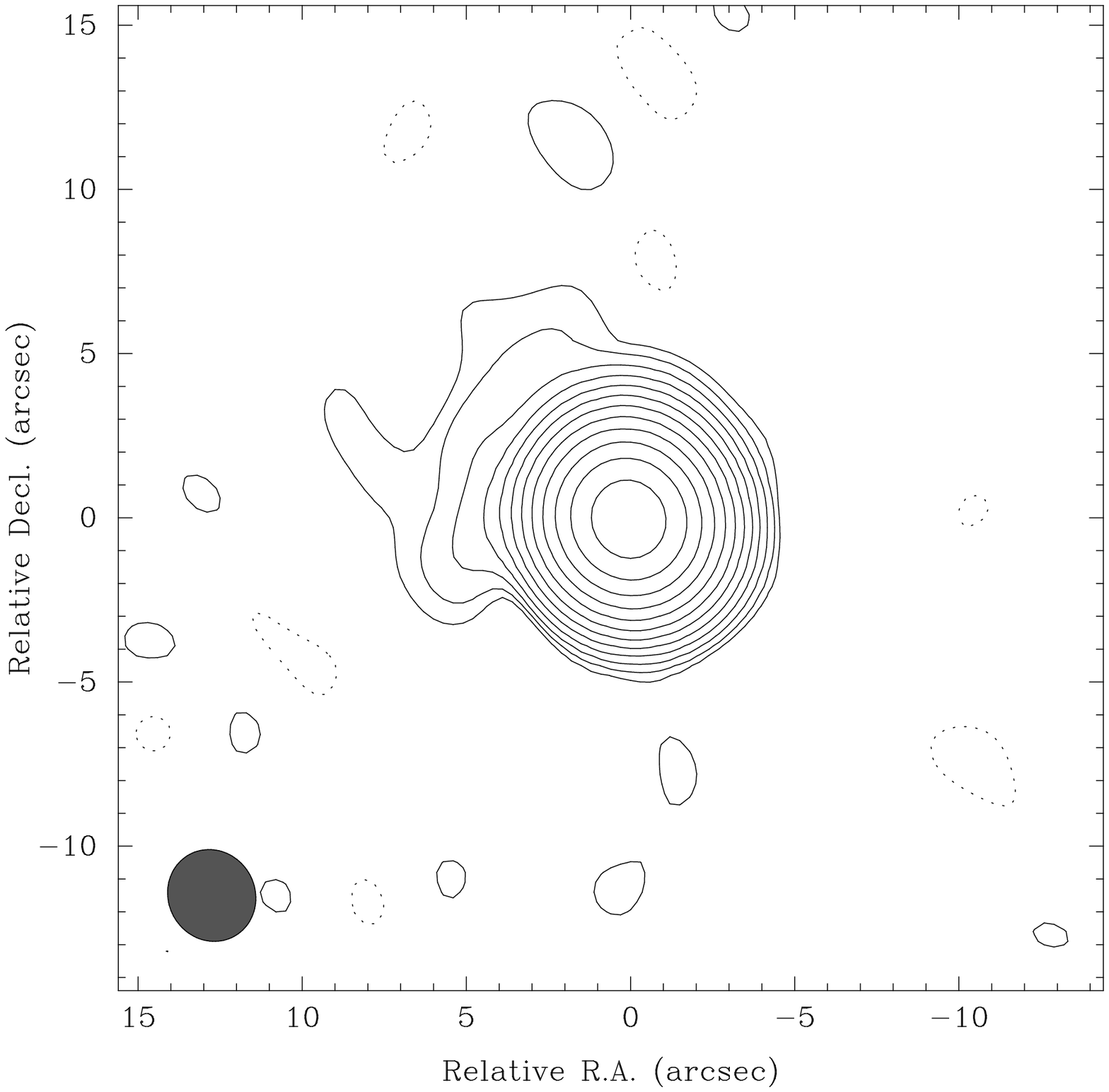]{A VLA A--array image of 2215+020 at
5.0\,GHz, made from an observation on February 26 1993.
The restoring beam (2.84$\times$2.64 arcseconds at ${\rm P.A.}=28\deg$)
is plotted in the lower left corner. Contour levels are drawn at
-0.03,0.03,0.06,...,61.4\% of the peak brightness of 0.606\,Jy/beam.
\label{fig5}}

\figcaption[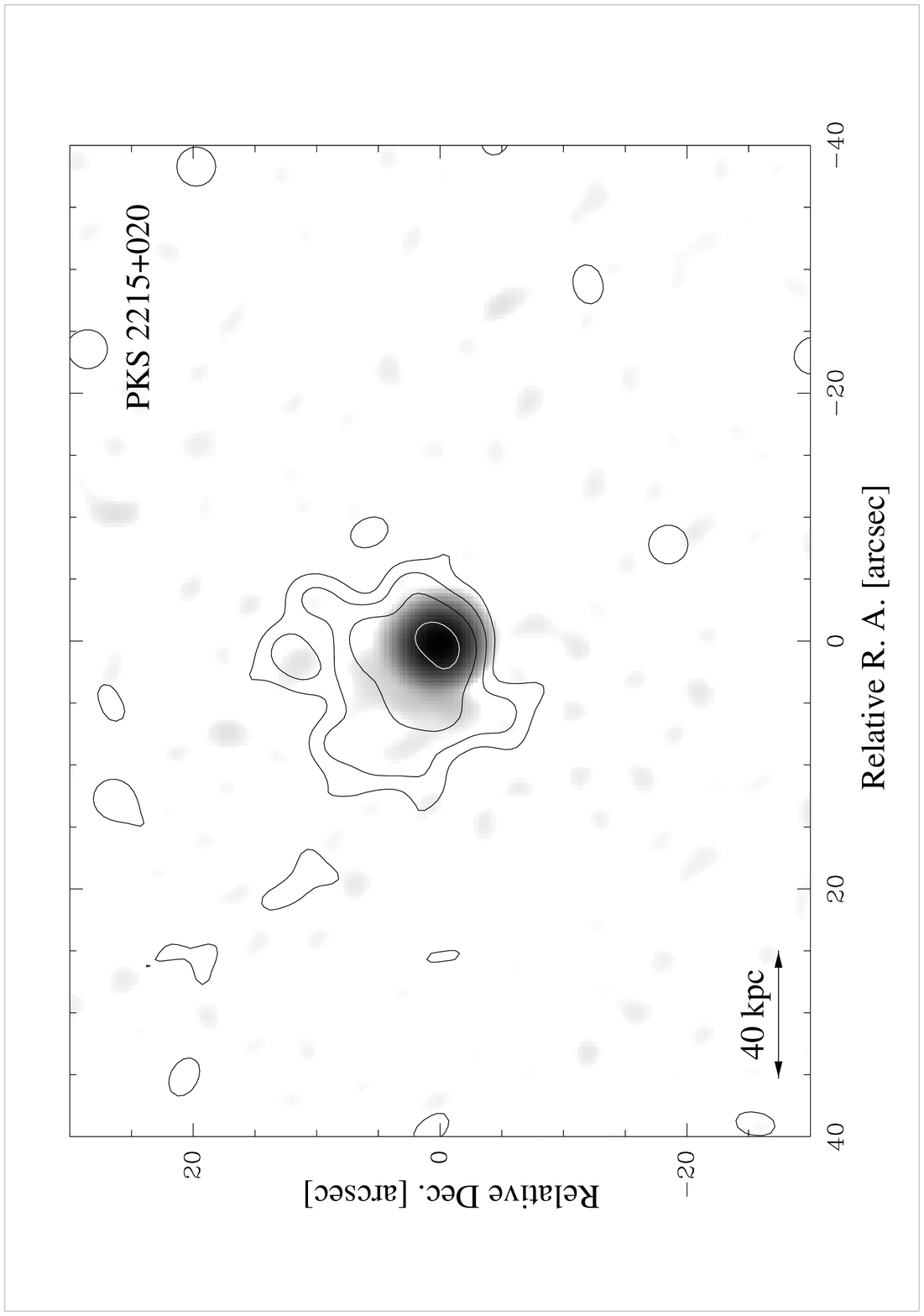]{ROSAT HRI image of 2215+020 (contours) 
overlaid on the VLA image (grey scale). The contour levels are drawn at
10, 20, 40, and 80\% of the peak of the X--ray emission. 
\label{fig6}}

\clearpage


\begin{table}[h]
\caption{Selected technical parameters of radio telescopes participated in
the VSOP observation of 2215$+$020.}
\label{tb:observation}
\medskip
\begin{center}
\begin{tabular}{lccccc} \hline\hline
  Telescope    & Code$^1$ & D$^2$ & T$_{\rm sys}^3$ & K$^4$ & SEFD$^5$ \\
                    &    & [m]  & [K] & [K/Jy] & [Jy]   \\ \hline\hline
HALCA Satellite     &    & 8    & 75  & 0.0043 & 17400 \\
Effelsberg, Germany & Eb & 100  & 40  & 1.50   & 30 \\
Medicina, Italy     & Mc & 32   & 70  & 0.10   & 700 \\
Noto, Italy         & Nt & 32   & 95  & 0.11   & 950 \\ 
Torun, Poland       & Tr & 32   & 50  & 0.13   & 380 \\
Green Bank, USA     & Gb & 43   & 30  & 0.29   & 90 \\ 
\multicolumn{6}{c}{VLBA, USA:}  \\
Saint Croix         & Sc & 25   & 28  & 0.082  & 310 \\ 
Hancock             & Hn & 25   & 33  & 0.080  & 330 \\ 
North Liberty       & Nl & 25   & 20  & 0.071  & 290 \\ 
Fort Davis          & Fd & 25   & 24  & 0.090  & 270 \\ 
Los Alamos          & La & 25   & 32  & 0.121  & 280 \\ 
Pietown             & Pt & 25   & 27  & 0.101  & 270 \\ 
Kitt Peak           & Kp & 25   & 31  & 0.099  & 310 \\ 
Owens Valley        & Ov & 25   & 34  & 0.100  & 340 \\ 
Brewster            & Br & 25   & 24  & 0 087  & 280 \\ 
Mauna Kea           & Mk & 25   & 34  & 0.097  & 340 \\\hline
\end{tabular}
\end{center}
\tablecomments{1 -- two--letter station code; 2 -- antenna diameter; 3 -- receiver
system temperature; 4 -- antenna sensitivity; 5 --  system equivalent flux
density.}
\end{table}

\begin{table}
\caption{Characteristics of the ground array and VSOP images}
\label{tb:images}
\footnotesize
\begin{center}
\medskip
\begin{tabular}{l|c|c}\hline\hline
& \multicolumn{2}{c}{Image:} \\  \cline{2-3}
& GVLBI & VSOP \\ \hline\hline
Beam size [mas] & 9.5$\times$2.8 & 4.5$\times$1.1 \\
Total CLEAN flux [mJy] & 451.3 & 465.8 \\
Map peak [mJy/beam] & 277.8 & 221.4 \\
Map noise [mJy/beam]     & 0.16 & 0.36 \\
Thermal noise [mJy/beam] & 0.08 & 0.07 \\
\hline
\end{tabular}
\end{center}
\end{table}

\begin{table}
\caption{Model fits of the structures in the ground array and VSOP images}
\label{tb:modelfit}
\footnotesize
\begin{center}
\medskip
\begin{tabular}{ccccccrcr}\hline\hline
    & & $S_{\rm peak}$     & $S$ & $r$ & $\phi$ & $d$ & $d_{\rm lim}$ & $T_{\rm b}$\\
    & & [mJy/beam]    & [mJy] & [mas] & [deg] & [mas] & [mas] & [K] \\
(1) & (2) & (3) & (4) & (5) & (6) & (7) & (8) & (9) \\ \hline
C:  & GVLBI& 278\z69  & 291\z100  &   0.0    & ...   & (0.9)    & 1.8 & $>1.9\times10^{11}$ \\
    & VSOP& 221\z81  & 300\z137  &   0.0   & ...    & 0.8\z0.3 & 0.8 & $9.1\times10^{11}$\\\hline
J7: & GVLBI& 4.2\z2.3 & 11.6\z6.8 & 3.3\z0.6 & 88\z11 &($<$0.1) & 2.3 & $>4.4\times10^9$    \\
    & VSOP& 3.7\z2.0 & 10.0\z5.8 & 3.0\z0.3 & 93\z6 &($<$0.1)  & 1.2 & $>1.5\times10^{10}$\\\hline
J6: & GVLBI& 2.6\z1.6 &  3.1\z2.5 & 7.8\z0.9 & 88\z6 &($<$0.1)  & 2.8 & $>8.1\times10^8$    \\
    & VSOP& 2.1\z1.1 &  2.8\z1.8 & 6.9\z0.4 & 84\z3 &($<$0.1)  & 1.5 & $>2.5\times10^9$ \\\hline
J5: & GVLBI& 2.5\z1.3 &  3.9\z2.4 &14.6\z0.4 & 90\z3 & (1.9)    & 2.7 & $>1.1\times10^9$    \\
    & VSOP& 2.5\z1.2 &  3.4\z2.0 &12.0\z0.4 & 89\z2 & 1.7\z0.8 & 1.4 & $2.0\times10^9$   \\\hline
J4: & GVLBI& 2.8\z1.4 & 12.1\z6.2 &23.7\z1.6 & 78\z4 & 6.2\z3.1 & 2.3 & $2.1\times10^8$     \\
    & VSOP&  \multicolumn{7}{c}{\it resolved}                        \\\hline
J3: & GVLBI& 1.6\z0.8 & 11.3\z5.7 &34.0\z1.0 & 91\z2 & 4.0\z2.0 & 2.3 & $5.0\times10^8$     \\
    & VSOP& 1.8\z0.8 &  9.4\z4.3 &32.1\z0.8 & 85\z2 & 3.8\z1.7 & 1.2 & $1.4\times10^9$  \\\hline
J2: & GVLBI& 9.1\z3.5 & 12.9\z6.1 &49.6\z0.6 & 84\z1 & 3.3\z1.3 & 2.3 & $7.4\times10^8$     \\
    & VSOP& 3.8\z1.7 & 12.0\z5.6 &49.0\z0.7 & 87\z1 & 3.1\z1.4 & 1.1 & $2.6\times10^9$     \\\hline
J1: & GVLBI& 18.0\z5.4& 110\z34   &59.5\z2.1 & 74\z2 &13.8\z4.1 & 1.9 & $4.2\times10^7$     \\
    & VSOP& 5.7\z2.8 & 123\z60   &57.1\z3.0 & 77\z2 &12.2\z6.0 & 0.9 & $1.6\times10^8$     \\\hline
\end{tabular}
\end{center}
\tablecomments{Column designation: 3 -- component peak brightness (from
the images); 4 -- component flux density; 5 -- distance from the core C; 6 --
position angle; 7 -- component size  (values in
brackets denote upper limits; listed sizes of $<$0.1\,mas denote
tentative values for the features in which the model fitting has
failed to produce a viable estimate); 8 -- limiting size (from
equation~1); 9 -- brightness temperature.}
\end{table}

\clearpage


\plotone{f1.eps}

\clearpage

\plotone{f2.eps}

\clearpage

\plotone{f3.eps}

\clearpage

\plotone{f4.eps}

\clearpage

\plotone{f5.eps}


\plotone{f6.eps}

\end{document}